\documentclass[twocolumn,aps,superscriptaddress]{revtex4-1}

\usepackage{graphicx}
\usepackage{amsmath, amsthm, amssymb}
\usepackage{epstopdf}
\usepackage{mathtools}
\usepackage{graphicx}
\usepackage{dcolumn}
\usepackage[tight]{subfigure}
\usepackage{amsmath}
\usepackage{verbatim}
\usepackage{inputenc}
\usepackage{color}
\usepackage{bm} 
\usepackage{natbib}
\usepackage{xspace}
\usepackage{mathtools}
\usepackage{hyperref} 
\usepackage{mathbbol}

\begin{document}
\preprint{APS/123-QED}
\title{On-Chip Cooling by Heating with Superconducting Tunnel Junctions}
\author{G. Marchegiani}
\email{giampiero.marchegiani@df.unipi.it} 
\affiliation{Dipartimento di Fisica dell'Universit\`{a} di Pisa, Largo Pontecorvo 3, I-56127 Pisa, Italy}
\affiliation{NEST Istituto Nanoscienze-CNR and Scuola Normale Superiore, I-56127 Pisa, Italy}

\author{P. Virtanen}
\email{pauli.virtanen@nano.cnr.it} 
\affiliation{NEST Istituto Nanoscienze-CNR and Scuola Normale Superiore, I-56127 Pisa, Italy}

\author{F. Giazotto}
\email{francesco.giazotto@sns.it} 
\affiliation{NEST Istituto Nanoscienze-CNR and Scuola Normale Superiore, I-56127 Pisa, Italy}

\date{\today}

\begin{abstract}
Heat management and refrigeration are key concepts for nanoscale devices operating at cryogenic temperatures.
The design of an on-chip mesoscopic refrigerator that works thanks to the input heat is presented, thus realizing a solid state implementation of the concept of \textit{cooling by heating}.
The system consists of a circuit featuring a thermoelectric element based on a ferromagnetic insulator-superconductor tunnel junction (N-FI-S) and a series of two normal metal-superconductor tunnel junctions (SINIS). The N-FI-S element converts the incoming heat in a thermovoltage, which is applied to the SINIS, thereby yielding cooling.
The cooler's performance is investigated as a function of the input heat current for different bath temperatures. We show that this system can efficiently employ the performance of SINIS refrigeration, with a substantial cooling of the normal metal island. Its scalability and simplicity in the design makes it a promising building block for low-temperature on-chip energy management applications.
\end{abstract}

\maketitle

\section{Introduction}
The unprecedented technological advancement running over the last two decades owns a great deal to the developments in nanotechnologies. The performance of electronic devices has been progressively increasing thanks to chip miniaturization, but this process is ultimately limited by the enormous heat production occurring at the nanoscale. 
The future for computation and telecommunications relies on \textit{quantum technologies}. Superconducting devices play a major role in this respect \cite{Clarke2008,Devoret2013}, but they require cryogenic temperatures. 

Since the beginning of the last century, it is possible to reach easily the temperature of 4.2 K, thanks to the use of liquid helium. The situation is different when sub-300 mK temperatures are required: in dilution fridges the refrigeration from 4.2 K to the base temperature $\sim10$ mK is realized through a multi-stage setup. These machines are bulky, expensive and still not so widespread in the world. This motivated an intense research activity on the development of on-chip cooler \cite{Giazotto2006,Muhonen2012} in order to replace the last stages of refrigeration. Their realization is also essential for instance for particle detection, making them appealing even for fundamental physics \cite{Enss2005}.

\begin{figure} [h!]
	\begin{centering}
		\includegraphics[width=0.5\textwidth]{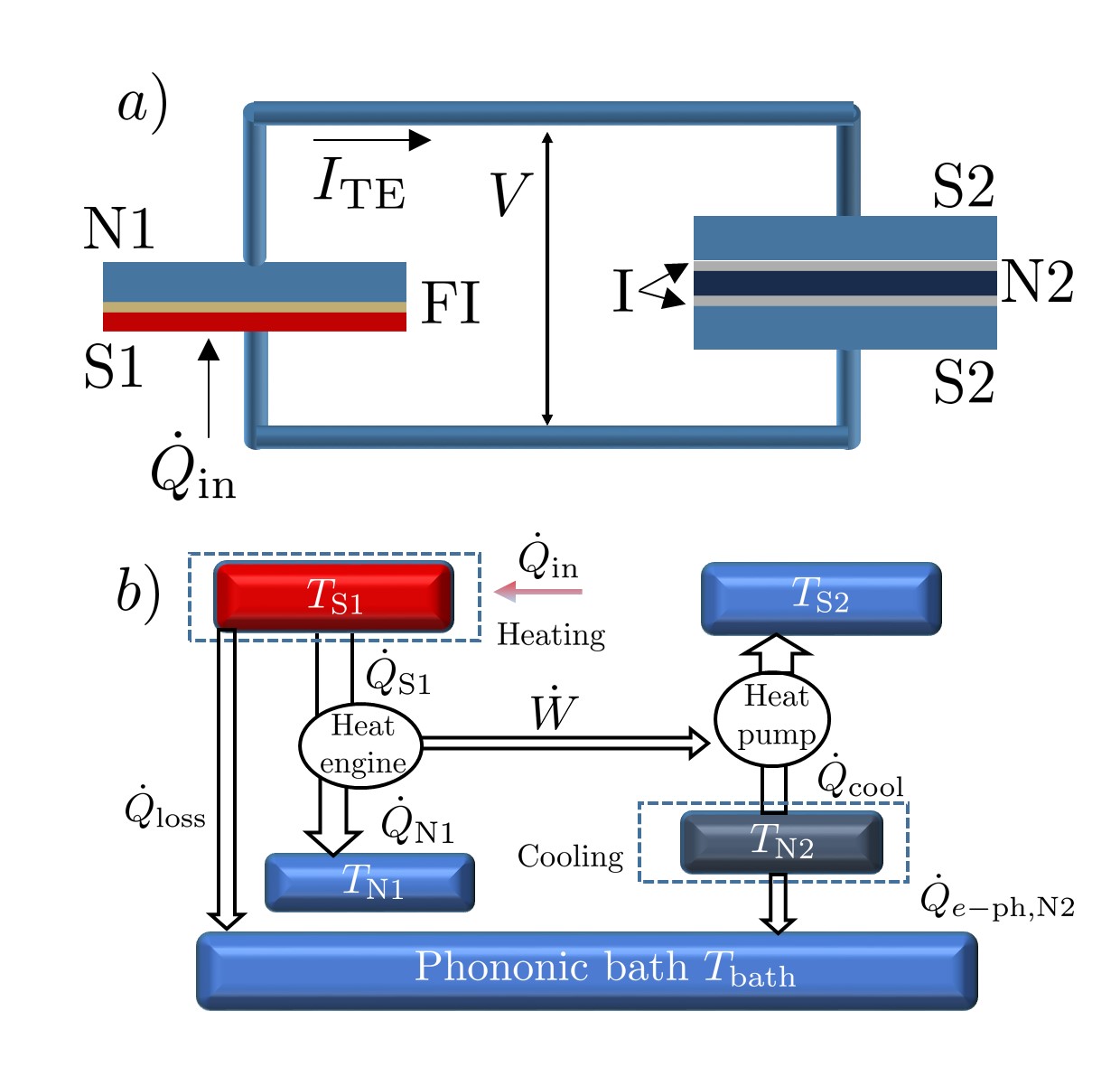}
		\caption{(color online). a) Scheme of the refrigerator. The N-FI-S element benefits from an input heat current $\dot Q_\textrm{in}$ to generate a thermocurrent $I_\textrm{TE}$, hence a voltage drop $V$ which is applied to the SINIS element, thus allowing cooling of its normal island. b) Thermodynamic scheme. The heat current from the hot reservoir $\dot Q_\textrm{S1}$(superconducting layer of the N-FI-S element) is partly converted into useful power $\dot W$ thanks to the thermoelectric element. This work allows for the extraction of heat from the normal side of the SINIS junction. 
		In the figure, heat exchange with the substrate phonons, which reside at temperature $T_\textrm{bath}$, is also shown.}
		\label{Fig1}
	\end{centering}
\end{figure}
Several solid-state based cooling concepts have been proposed and investigated over the years \cite{Giazotto2006,Muhonen2012}.
Among them, the normal metal-insulator-superconductor (NIS) tunnel junction refrigeration provides a powerful method to cool down the electronic temperature of the normal metal \cite{Nahum1994,Bardas1995}. Its operation is based on the presence of the superconducting energy gap \cite{Tinkham2004}, which provides an energy filter allowing the transmission of only hot quasiparticles, and yielding maximum cooling, when the voltage across the junction is just below the gap.
These structures have been intensively investigated \cite{Lowell2013,ONeil2012,Vasenko2010} and their performance can be improved by connecting two junctions in series back-to-back so to realize a SINIS structure\cite{Leivo1996,Pekola2004}. Their operation requires a voltage bias, usually controlled externally.

Here we present the design and quantitative analysis of a mesoscopic cooler that i) works thanks to heat extracted from the environment,
ii) does not require external electrical bias, iii) is feasible to experimental implementation with current state-of-the-art nanotechnology. The system benefits from a thermoelectric element of recent prediction \cite{Ozaeta2014,Machon2013} and discover \cite{Kolenda2016,Kolenda2017} in spin split and filtered superconducting tunnel junction to produces an effective voltage that leads to cooling, thanks to SINIS energy filtering. Since the system is only fueled by heat, this mechanism can be described as \emph{cooling by heating}\cite{Mari2012,Clauren2012}. 

In the following we analyze the performance and the limitations of this system. We shows that, with a proper tuning of the main parameters (such as the superconducting energy gap), this scheme is capable of leveraging quite efficiently the potential of SINIS refrigeration.

\section{Model and numerical analysis}
A schematic representation of the cooler is shown in the panel a) of Fig. \ref{Fig1}. 
The cooler consists of two main elements: a normal metal-ferromagnetic insulator-superconductor tunnel junction (N-FI-S), which acts as a thermoelectric element and a SINIS structure, the effective refrigerator. The two elements are connected through a superconducting wire. Throughout the paper we neglect the heat exchange between the S layer (i.e the hot layer, as discussed below) of the N-FI-S element and the superconducting wire, which could be obtained, for instance, by using a superconductor with larger gap for the wire. The cooler works as follows. The heat current $\dot Q_\textrm{in}$ harvested by the system raises the electronic temperature of one side of the thermoelectric element. As a consequence, a thermocurrent $I_\textrm{TE}$ flows in the circuit \cite{Giazotto2015,Ozaeta2014} and a voltage $V$  develops across the SINIS element. For particular values of the parameters, this configuration yields cooling of electrons, lowering the electronic temperature of the N island of the SINIS with respect to the temperature of the substrate phonons. The thermodynamical scheme is portrayed in Fig. \ref{Fig1}, panel b). The system features a heat engine, i.e. the N-FI-S element\cite{MarchegianiEngine}, coupled to a heat pump, the SINIS. Hence the absorbed heat is first partly converted into useful work by the N-FI-S junction.  Then this amount of work is used to extract heat from the normal layer of the SINIS. In the same figure, we indicate the main losses and heat exchanges in the system, which are discussed below.

In the following, we provide a standard tunneling description of electrical and thermal transport through the junctions. For later convenience we define here the quasiparticle BCS\cite{Tinkham2004} density of states (DOS) $N_i(E)=\left|{\mathrm{Re}\left[\frac{E+j\Gamma_i}{\sqrt{(E+j\Gamma_i)^2-\Delta_i^2}}\right]}\right|$ 
where $j$ is the complex unity, $i=1,2$ refers respectively to the N-FI-S junction and each N-I-S junction of the SINIS, which we assume perfectly symmetric for simplicity.
Here E is the electron's energy with respect to the chemical potential of the superconductor, $\Gamma_i$ is the phenomenological broadening parameter \cite{Dynes1984} and $\Delta_i(T)$ is the superconducting pairing potential \cite{Tinkham2004}. Similarly we introduce the anomalous function $F(E)=\mathrm{sgn}(E)\left|\mathrm{Re}\left[\frac{\Delta_1}
{\sqrt{(E+j\Gamma_1)^2-\Delta_1^2}}\right] \right|$. We also assume that the quasiparticle distribution at temperature $T$ is always given by the Fermi-Dirac distribution $f(V,T)=\{1+\exp[(E+eV)/k_\textrm B T]\}^{-1}$, where $-e$ is the electron charge and $k_\textrm B$ is the Boltzmann constant. Throughout this paper we refer to the expression $f(0,T)$ as $f(T)$.

In order to give a self-contained exposition, we first describe the two elements separately.
We start with the N-FI-S thermoelectric element. In a conductor the dominant contribution to thermoelectricity comes from particle-hole asymmetry in the energy DOS of quasiparticles \cite{Mermin1978}. In this structure, this condition is fulfilled by exploiting two separate effects i) the Zeeman splitting in the presence of an exchange magnetic field, which breaks the particle-hole symmetry for each spin band; ii) the spin filtering of the junction.
Both effects are provided by the ferromagnetic insulator layer. On the one hand it generates an exchange contribution $h_\textrm{exc}$ as a result of the interaction of its dipole with the electrons spin of the superconductors, on the other hand it acts as a natural spin filter with polarization $P=(G_\uparrow-G_\downarrow)/(G_\uparrow+G_\downarrow)$ \cite{Moodera2007}, where $G_{\uparrow,\downarrow}$ is the spin up (down) junction normal state conductance.

More precisely, we indicate with $N_{\uparrow,\downarrow}$ the DOS of the spin up ($\uparrow$) and spin down ($\downarrow$) bands. In the presence of an homogeneous exchange field in the superconductor S we have $N_{\uparrow(\downarrow)}=N_1(E\pm h_{\textrm{exc}})/2$.
This approximation holds for ultrathin layers, where the thickness is smaller than the superconducting decay length of the exchange interaction. The superconducting pairing potential  depends also on the exchange interaction  $\Delta_1(h_{\textrm{exc}},T_\textrm{S1})$ and can be determined through a standard self consistent calculation \cite{Giazotto2015}.

The quasiparticle DC current through the N-FI-S element is given by \cite{Giazotto2006,Ozaeta2014}
\begin{equation}
I_\textrm{TE}\text=\frac{1}{e R_1} \int_{-\infty}^{+\infty}dE\mathcal N(E)[f(T_\textrm{S1})-f(V,T_\textrm{N1})]
\label{eq:thermocurrent}
\end{equation}
where $R_1$ is the normal-state resistance of the junction and $V$ is the bias voltage.
Here $\mathcal N=N_+ + PN_-$, with $N_{\pm}=N_\uparrow \pm N_\downarrow$ and $T_\textrm{S1(N1)}$ are respectively the temperatures of the S(N) layers. 

\begin{figure}[tp]
\begin{centering}
\includegraphics[width=0.5\textwidth]{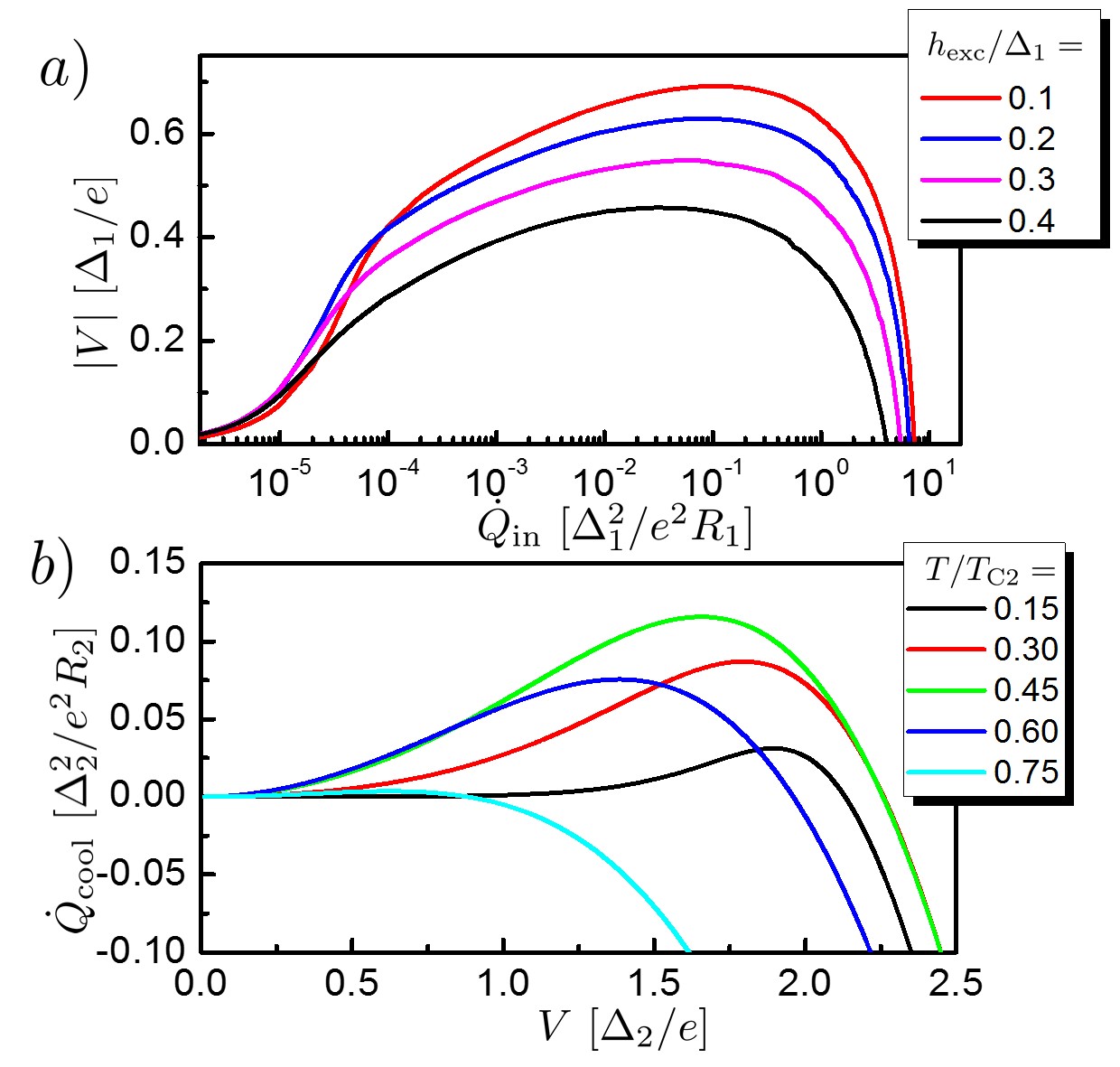}
\caption{(color online). a) Normalized absolute thermovoltage provided by the thermoelectric element vs normalized input heat power at open-circuit for different values of the exchange field $h_\textrm{exc}$ at $T_\textrm{bath}=0.3$ K. For the S layer, we consider an ultrathin film of Al, with critical temperatures equal to $T_\textrm{C1}=3\ \textrm K$, hence $\Delta_1=1.764\ k_\textrm B T_\textrm{C1}\simeq456\ \mu$eV. The electron phonon coupling constant is  $\Sigma_\textrm{S1}=\Sigma_\textrm{Al}=0.2\times 10^9$ W K$^{-5}$ m$^{-3}$, the volume is $\mathcal V=3.0\times10^{-17}$ m$^3$ and the Dynes parameter is $\Gamma_1=10^{-4} \Delta_1$. The tunnel junction parameters are $R_1=1\ \Omega$ and $P=0.95$. 
b) Normalized cooling power vs normalized bias voltage for a SINIS refrigerator at different bath temperatures for $\Gamma_2=10^{-4} \Delta_2$.}
\label{Fig2}
\end{centering}
\end{figure}
\subsection{Open Circuit Thermovoltage}
In the open-circuit configuration, i.e. for $I_\textrm{TE}=0$, the thermoelectric effect manifests itself as a Seeback voltage across the N-FI-S element. Throughout this paper, we consider a configuration where the heat is provided to the superconducting layer of the thermoelectric element. This choice is motivated by two reasons: i) the power losses, which are related to the electron-phonon interaction, are exponentially suppressed in the superconducting state, ii) the thermoelectric effect is stronger when the superconducting side is heated \cite{Giazotto2015}.
By assuming a constant heat current $\dot{Q}_\textrm{in}$ fueling the system, the thermovoltage is obtained by solving the following system of equation self-consistently for the unknowns $T_\textrm{S1},V$
\begin{align}
I_\textrm{TE}=0\quad,\quad \dot{Q}_\textrm{in}=\dot{Q}_\textrm {S1}+\dot{Q}_\textrm{loss}
\,.
\label{eq:opencircuit}
\end{align}
Here $T_\textrm {S1}$, $\dot{Q}_\textrm {S1}$ are the temperature and the heat current from the S layer of the thermoelectric element, respectively, with 
\begin{equation} 
\dot{Q}_\textrm {S1}=\frac{1}{e^2 R_1}\int_{-\infty}^{+\infty}dE E\mathcal N(E)[f(T_\textrm {S1})-f(V,T_\textrm {N1})].
\end{equation} 
According to the second principle of thermodynamics, this amount of power is only partially converted in useful power $\dot W=-I_{TE} V$ and the remaining part ($\dot Q_\textrm{N1}$) is released to the normal layer of the thermoelectric element
\begin{equation}
\dot{Q}_\textrm{N1}=\frac{1}{e^2 R_1}\int_{-\infty}^{+\infty}dE(E+eV)\mathcal N(E)[f(T_\textrm{S1})-f(V,T_\textrm{N1})]
\end{equation} 
as depicted in Fig. \ref{Fig1}, panel b). 
The losses due to electron-phonon interaction are accounted for through the term $\dot{Q}_\textrm{loss}$: in a spin-split superconductor this interaction is described by \cite{Bergeret2017}
\begin{equation}
\dot{Q}_\textrm{e-ph}=\frac{-\Sigma_\textrm{S1}\mathcal{V}_\textrm{S1}}
{96 \zeta(5)k_\textrm B^5}\iint_{-\infty}^{+\infty} dE d\varepsilon E
\varepsilon^2 \mathrm{sgn}(\varepsilon) M(E,E+\varepsilon)I(E,\varepsilon)
\end{equation}
where
\begin{equation}
\nonumber
I(E,\varepsilon)=\coth\frac{\varepsilon}{2k_BT_\textrm{bath}}(f(E)-f(E+\varepsilon))
-f(E)f(E+\varepsilon)+1
\end{equation}
and
\begin{equation}
\nonumber
M(E,E')=2\sum_{\sigma=\uparrow,\downarrow} [N_\sigma(E)N_\sigma(E')
-F_\sigma(E)F_\sigma(E')]
\end{equation}
with
$F_{\uparrow,\downarrow}(E)=F(E\pm h_\textrm{exc})/2$. Above, $\Sigma_\textrm{S1}$, $\mathcal V_\textrm{S1}$ are the electron-phonon coupling constant and the volume of the superconducting layer of the N-FI-S element, respectively, and $\zeta$ is the Riemann Zeta function. In this model the phonons of all the elements are assumed to reside at the temperature of the substrate phonons $T_\textrm{bath}$, because for thin layers the Kapitza resistance is negligible \cite{Giazotto2006}. Furthermore we assume that the normal side of the thermoelectric element is well thermalized with substrate phonons, i.e. $T_\textrm{N1}=T_\textrm{bath}$. 

In the first panel of Fig. \ref{Fig2} we plot the absolute thermovoltage $|V|$ as a function of $\dot{Q}_\textrm{in}$ for different values of the exchange field $h_{\textrm{exc}}$. 
The plot summarizes the main features of the thermoelectric effect.
i) The thermovoltage increases abruptly for small input heat current $\dot{Q}_\textrm{in}\gtrsim 10^{-5} \Delta_1^2/e^2R_1$, where $\Delta_1$ is the zero field and zero temperature pairing potential of the S layer. This fact is somewhat related to the particular choice for the subgap conductance $\Gamma_1/\Delta_1=10^{-4}$. ii) The thermovoltage displays a non monotonic dependence on $\dot{Q}_\textrm{in}$: the power decreases to zero for $\dot{Q}_\textrm{in}\gtrsim  \Delta_1^2/e^2R_1$ after reaching a maximum. This behaviour is related to the existence of the superconducting critical temperature $T_\textrm{C1}$ of the S layer: No thermoelectric effect is possible when the S layer is in the normal state \cite{Ozaeta2014}.
iii) The thermovoltage is non-monotonic with the exchange field.
iv) The thermovoltage generated is always smaller than $\Delta_1/e$. 
In the calculation, we consider proper materials for the N-FI-S junction: europium chalcogenides (as EuO, EuS or EuSe) for the FI coupled to ultrathin films of Al for the superconducting layer, where large polarizations $P$ up to nearly 100\%  have been well established \cite{Moodera1988,Moodera1993,Hao1990,Strambini2017}.
\subsection{SINIS refrigeration}
Here we briefly discuss the SINIS refrigeration process. The cooling power $\dot{Q}_\textrm{cool}$, i.e. the heat current leaving the normal island of the SINIS element, is an even function of the voltage bias $V$, namely\cite{Giazotto2006,Muhonen2012}
\begin{equation}
\footnotesize
\dot{Q}_\textrm{cool}=\frac{2}{e^2 R_2}\int_{-\infty}^{+\infty}dE\left(E+\frac{eV}{2}\right) N_{2}(E)\left[f\left(\frac{V}{2},T_\textrm{N2}\right)-f(T_\textrm{S2})\right]
\label{eq:Qcool}
\end{equation}
where the factor two comes from the presence of two NIS junctions of normal state resistance $R_2$. Here $T_\textrm{N2(S2)}$ are the temperatures of the S and N layer of the SINIS, and the pairing potential $\Delta_2(T_\textrm{S2})$ is only a function of the temperature, since no effective exchange field is present in the SINIS.

Panel b) of Fig. \ref{Fig2} shows the cooling power against the voltage bias for different bath temperatures $T_\textrm{bath}$, with $T_\textrm{S2}=T_\textrm{N2}=T_\textrm{bath}$. Two main features can be captured by the plot:
i) for a given $T_\textrm{bath}$ there is an optimal voltage bias for refrigeration. In the low temperature limit  it is given by $eV\sim 2(\Delta_2-0.66 k_\textrm B T_\textrm{bath})$ \cite{Giazotto2006}, where $\Delta_2$ is the zero temperature superconducting pairing potential. When the voltage is well above the gap, no refrigeration is possible. 
ii) There is an optimal temperature for refrigeration, $T_\textrm{bath}\simeq 0.45\ T_\textrm{C2}$ where $T_\textrm{C2}$ is the critical temperature of the superconductor of the SINIS element. The cooling power is small at low temperature and vanishes when $T_\textrm{bath}$ approaches $T_\textrm{C2}$. 
This last feature can be explained as follows: at low temperatures a small number of quasiparticles is available for refrigeration, whereas at high temperature the closing of the superconducting pairing potential destroys the effect. In a NIN junction a voltage bias always produces heating (i.e, Joule heating). 

These plots show the fundamental operation of the refrigerator: The optimal working point of the cooler is achieved when the thermoelectric element provides a voltage of order $V\sim 2\Delta_2/e$. Since the thermovoltage at open circuit is always $\leq\Delta_1/e$ the superconducting pairing potential of the N-FI-S element should be bigger than the corresponding gap of the superconductor in the SINIS element, to obtain the best performance.

\subsection{Closed-circuit operation}
When the N-FI-S element is connected to the SINIS refrigerator a finite current $I_\textrm{TE}$ flows in the circuit, and the resulting thermovoltage $V$ is then applied across the SINIS element. In the modeling of this system, the first of Eqs. \ref{eq:opencircuit} must be replaced with the current conservation in the circuit, namely:
\begin{align}
I_\textrm{TE}=I_\textrm{NIS}\quad,\quad \dot{Q}_\textrm{in}=\dot{Q}_\textrm{S1}+ \dot{Q}_\textrm{loss}
\,
\label{eq:closecircuit}
\end{align}
where 
\begin{equation}
I_\textrm{NIS}=\frac{1}{e R_2}\int_{-\infty}^{+\infty}dE N_2(E)[f(-V/2,T_\textrm{N2})-f(T_\textrm{S2})]
\end{equation} is the quasiparticle current through the SINIS element.
The picture is now less intuitive since the voltage $V$ depends also on the cooler parameters: the temperatures $T_\textrm{N2},T_\textrm{S2}$, the gap $\Delta_2$ and the junction normal state resistance $R_2$. 
 
\begin{figure}[tp]
	\begin{centering}
		\includegraphics[width=0.5\textwidth]{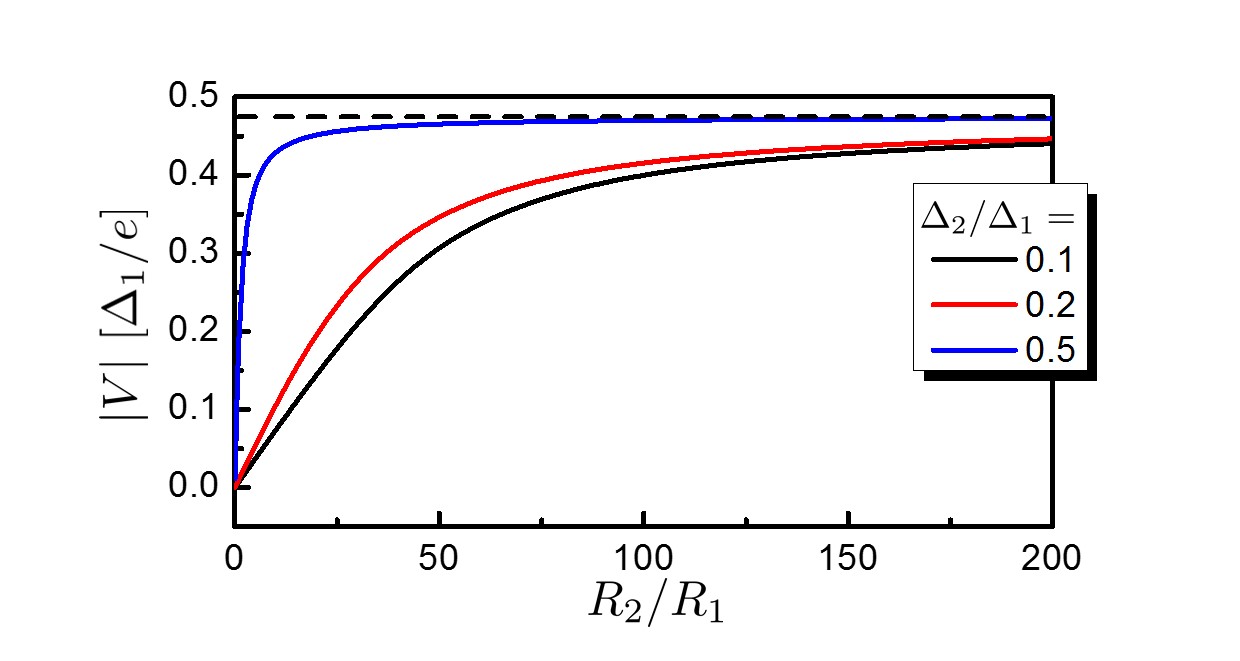}
		\caption{(color online). Absolute thermovoltage bias vs normal state resistance of the NIS junction for different values of $\Delta_2/\Delta_1$. Parameters are $\dot Q_\textrm{in}=2$ nW, $R_1=1\ \Omega$, $P=0.95$, $T_\textrm{C1}=3.0$ K, $\Gamma_1= 10^{-4}\ \Delta_1$, $\Gamma_2= 10^{-4}\ \Delta_2$ and $h_{\textrm{exc}}=0.1\ \Delta_1$. The open-circuit voltage at the same input power is drawn as dashed line.}
		\label{Fig3}
	\end{centering}
\end{figure}

We investigate this feature in Fig. \ref{Fig3}, where we plot the thermovoltage as a function of $R_2$ for some values of the ratio $\Delta_2/\Delta_1$. We consider a specific case, i.e. $\dot{Q}_\textrm{in}=0.1 \Delta_1^2/e^2R_1\simeq2$nW and $T_\textrm{bath}=0.3 $K but the results are similar quite generally. As before, the layers of the SINIS element are kept at the bath temperature $T_\textrm{S2}=T_\textrm{N2}=T_\textrm{bath}$. The thermovoltage increases monotonically with the ratio $R_2/R_1$ and reaches a plateau for values $R_2/R_1\gtrsim 100$. The asymptotic value increases with the ratio $\Delta_2/\Delta_1$ and approaches the open-circuit value, depicted as a dashed line. Note that even for $\Delta_2/\Delta_1=0.1$, the difference with the open-circuit value is smaller than 10\%. Hence in the high impedance limit the thermoelectric element operation is practically decoupled from the SINIS refrigeration and the open-circuit value still represents a useful approximation. This makes it a good operating point: The cooler's performance is then optimized by considering a proper ratio between the gap parameters to match the condition $V\simeq2\ \Delta_2/e$. 

\subsection{Cooling Power}
In view of a realistic implementation for sub-kelvin refrigeration, we choose Cu for the normal metal and Al as superconductor of the SINIS element, with $\Delta_2\simeq\ 200\ \mu \textrm{eV}$ and critical temperature $T_\textrm{C2}=\Delta_2/1.764 k_\textrm B \simeq 1.315\ \textrm K$. This setting guarantees a ratio $\Delta_1/\Delta_2\simeq2.3$ in order to produce a proper voltage for the maximum thermovoltage at exchange field $h_{\textrm{exc}}=0.1\ \Delta_1$. We use for the normal state resistance of each junction of the SINIS the value  $R_2=100\ \Omega$. In Fig. \ref{Fig4}, panel a) we display the cooling power $\dot{Q}_\textrm{cool}$ as a function of the input power $\dot{Q}_\textrm{in}$ for different bath temperatures. The electrodes of the SINIS are kept at bath temperature $T_\textrm{N2}=T_\textrm{S2}=T_\textrm{bath}$. 

The cooling power increases abruptly with the input power around $10$ pW, accordingly to the open-circuit dependence of the voltage (see Fig. \ref{Fig2} panel a). 
The cooling power displays a maximum for input power in the range $5-60\ \textrm{nW}$ and then goes to zero for powers of the order $1000\ \textrm{nW}$. This non monotonic behaviour reflects the open-circuit voltage dependence (see Fig. \ref{Fig2} panel a), since the thermoelectric effects holds only when the S layer of the N-FI-S stays in the superconducting state. Note also that the cooling power is maximum at $T_\textrm{bath}=0.5\ \textrm K$. This feature is related to the nature of the SINIS, which admits an optimal working temperature $T_\textrm{bath}\simeq 0.45\ T_\textrm{C2}\sim 0.58\ \textrm K$ (see Fig. \ref{Fig2} b). The cooling power is in the range of $1-30\ \textrm{pW}$ for the values chosen for the numerical plot. The maximum cooling power under optimal conditions in terms of temperature and voltage biasing depends only on the SINIS: in particular on the value of the pairing potential $\Delta_2$ and the normal state resistance of the junction $R_2$. Typically, improved performance can be obtained by decreasing the normal state resistance $R_2$ of each NIS junction, since the cooling power scales inversely with it (see Eq. \ref{eq:Qcool}). In this system this trend is limited by the normal state resistance of the N-FI-S junction $R_1$, because the thermovoltage decreases when the two resistances are of the same order (see Fig. \ref{Fig3}), and the optimal biasing is therefore lost. This behaviour is shown in Fig. \ref{Fig5}, where the cooling power is plotted against the normal state resistance of each NIS junction, for a nearly optimal input power $\dot Q_\textrm{in}=5$ nW. As one can see, the maximum cooling power is roughly twenty times bigger than the maximum of Fig. \ref{Fig4} a).
\subsection{Efficiency}
\begin{figure}[tp]
	\begin{centering}
		\includegraphics[width=0.5\textwidth]{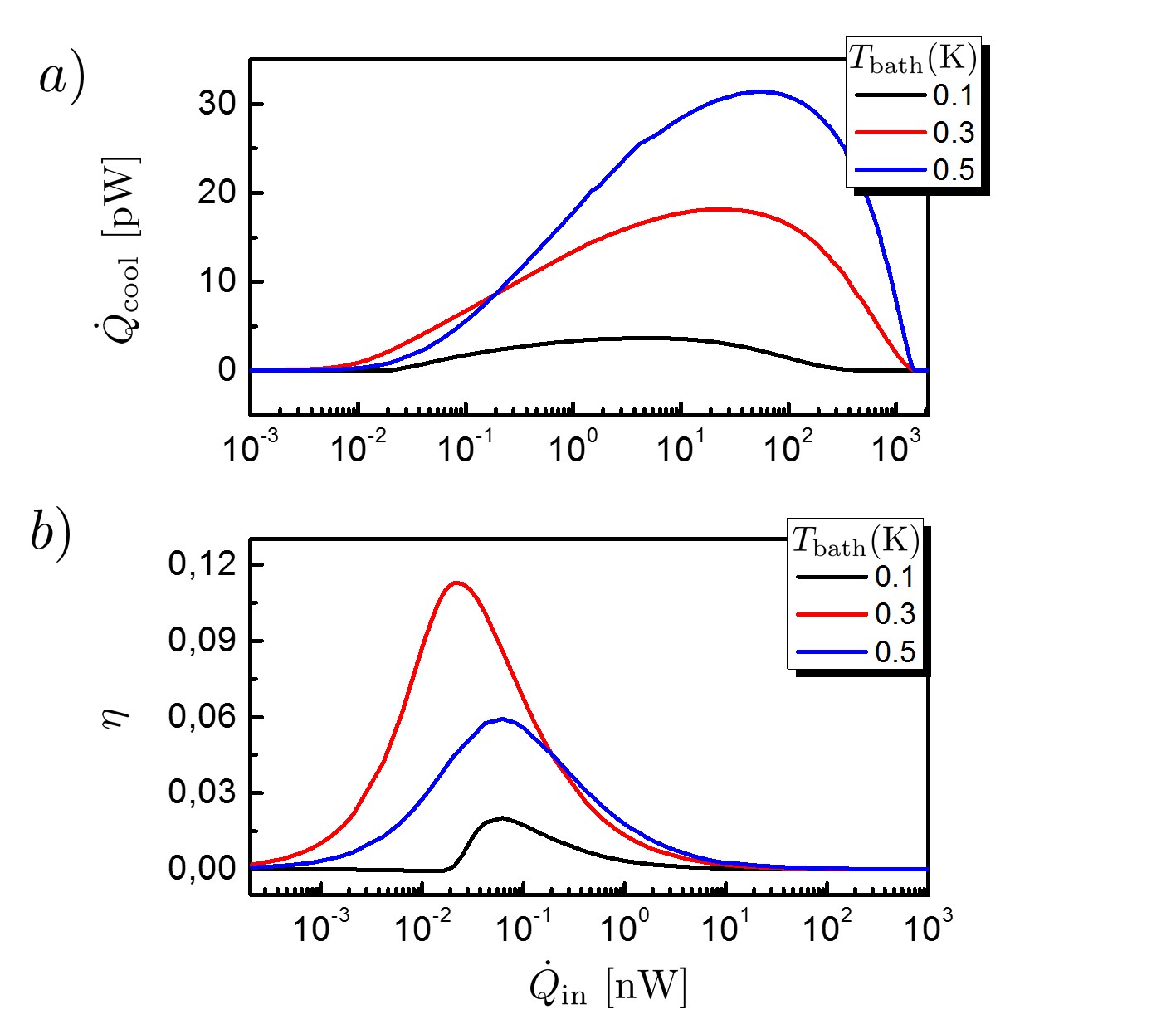}
		\caption{(color online). (a) Cooling power  and (b) efficiency vs input power for different bath temperatures, heating the superconductor layer of the thermoelectric N-FI-S element. Parameters: $R_2=100\ \Omega$, $T_\textrm{C2}=1.315\ \textrm K$, $T_\textrm{N2}=T_\textrm{S2}=T_\textrm{bath}$ and $\Gamma_2=10^{-4}\Delta_2$ for the SINIS element. For the thermoelectric element we set $P=0.95$, $R_1=1\ \Omega$, $\Gamma_1=10^{-4}\Delta_1$ where $\Delta_1=456\ \mu\textrm{eV}$ and $h_{\textrm{exc}}=0.1\ \Delta_1$. }
		\label{Fig4}
	\end{centering}
\end{figure}

\begin{figure}[tp]
	\begin{centering}
		\includegraphics[width=0.5\textwidth]{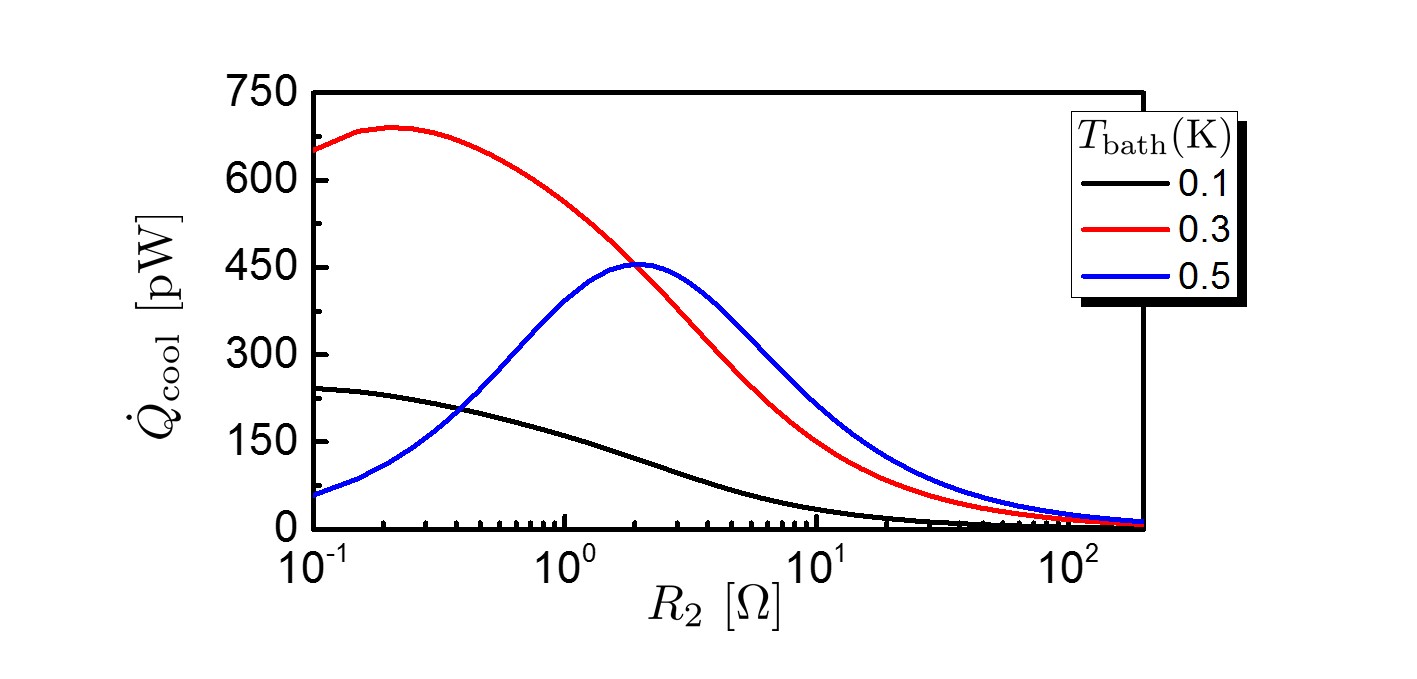}
		\caption{(color online). (a) Cooling power vs NIS normal state resistance for different bath temperatures, heating the superconductor layer of the thermoelectric N-FI-S element. Parameters: $T_\textrm{C2}=1.315\ \textrm K$, $T_\textrm{N2}=T_\textrm{S2}=T_\textrm{bath}$ and $\Gamma_2=10^{-4}\Delta_2$ for the SINIS element. For the thermoelectric element we set $P=0.95$, $R_1=1\ \Omega$, $\Gamma_1=10^{-4}\Delta_1$ where $\Delta_1=456\ \mu\textrm{eV}$ and $h_{\textrm{exc}}=0.1\ \Delta_1$. }
		\label{Fig5}
	\end{centering}
\end{figure}

In Fig. \ref{Fig4} b) the efficiency $\eta$ is plotted against the input power $\dot{Q}_\textrm{in}$, for the same parameters used in Fig. \ref{Fig4} a). We define this efficiency as the ratio between the cooling power and the input power $\eta=\dot{Q}_\textrm{cool}/\dot{Q}_\textrm{in}$. Note that, by definition, this quantity does not need to be smaller than one. The efficiency in the absence of losses, which generally lower its value, is given, from a fundamental thermodynamical point of view, by the product between the heat engine efficiency $\eta_\textrm{HE}$ and the coefficient of performance (COP) of the SINIS refrigerator. Hence the thermodynamical limit is \cite{Benenti2016}
\begin{equation}
\eta_\textrm{Max}= \eta_\textrm{HE}\cdot\textrm {COP}\leq \left(1-\frac{T_\textrm{bath}}{T_\textrm{S1}}\right)\left(\frac{T_\textrm{S2}}{T_\textrm{N2}}-1\right)^{-1}.
\end{equation}
This value diverges for $T_\textrm{S2}=T_\textrm{N2}$, which is the situation considered in the figures. 
The efficiency $\eta$ displays a bell shape behaviour: it reaches a maximum for input power in the range $10-60$ pW, in correspondence of the abrupt change in the cooling power displayed before, and then decreases rapidly to zero with some power law. 
Note that in this configuration the COP is always smaller than 1, with a maximum of $\eta\sim 0.1$ at $T_\textrm{bath}$=300 mK: this is partly related to the specific values chosen for $R_2, R_1$ and $\Delta_2$. On the one hand, at fixed input power $\dot Q_\textrm{in}$, its value depends only on the cooling power (hence on $\Delta_2, R_2$). On the other hand, the input power required for an optimal bias of the SINIS refrigerator depends on the normal state resistance $R_1$ of the thermoelectric junction: in particular this power decreases by increasing $R_1$  (for our parameters, at $T_\textrm{bath}=0.3$ K  this power is around 0.1 $\Delta_1^2/e^2 R_T$, as shown in Fig.\ref{Fig1} panel a)). Although is theoretically possible to find parameters with $\eta>1$, this case is not achieved, especially in the good operating regions in terms of cooling power. It is straightforward to see that the efficiency scales proportionally to $(\Delta_2/\Delta_1)^2R_1/R_2$. However, either by decreasing the gap ratio $\Delta_1/\Delta_2$ or by increasing the ratio $R_1/R_2$ in order to increase the efficiency, the voltage generated decreases and the optimal voltage bias is lost (see Fig. \ref{Fig3}). In Fig. \ref{Fig4} b) the maximum cooling power region the efficiency is of order $10^{-3}$, whereas in the optimized situation describe in Fig. \ref{Fig5}, $\eta$ at maximum power is roughly $700$ pW/5 nW$\simeq0.14$ (the efficiency follows exactly the same dependence of the cooling power, since the input power $\dot Q_\textrm{in}$ is fixed in the figure). More generally, it is possible to show that the efficiency at optimal bias in terms of maximum power for a NIS refrigerator alone is around 0.25 \cite{Giazotto2006}.
 
\subsection{Steady-state electronic Temperature}
Finally we consider the cooling of the normal island of the SINIS element. The thermal model of the normal island is sketched in Fig. \ref{Fig1}a. We assume that the heat coming out of the normal layer does not change the electronic temperature $T_\textrm{S2}$ of the superconducting layers, which are assumed to remain at $T_\textrm{bath}$. This can be realized experimentally through the use of quasiparticle traps which thermalize the superconductor \cite{Agulo2004,Rajauria2012,Court2008}. Moreover, $\dot{Q}_{e\textrm{-ph,N2}}=\Sigma_\textrm{N2} \mathcal V_\textrm{N2} (T_\textrm{bath}^n-T_\textrm{N2}^n)$,
where $\Sigma_\textrm{N2}$ is the electron phonon coupling constant of the specific metal, $\mathcal V_\textrm{N2}$ is the volume of the normal island and the exponent $n$ is characteristic of the material. For Cu we have $n=5$ and $\Sigma_\textrm{N2}=\Sigma_\textrm{Cu}=2.0\times 10^9\ \textrm W \textrm K^{-5} \textrm m^3$ \cite{Giazotto2006}.  
The system of Eqs. \ref{eq:closecircuit} must be supplemented with the heat balance equation
\begin{equation} 
\dot Q_{e\textrm{-ph,N2}}(T_\textrm{N2},T_\textrm{bath})=\dot{Q}_\textrm{cool}(T_\textrm{N2},T_\textrm{bath}),
\label{eq:heatbalance}
\end{equation}
yielding a system of the three equations 
which we solve numerically. 
\begin{figure}
	\begin{centering}
		\includegraphics[width=0.4\textwidth]{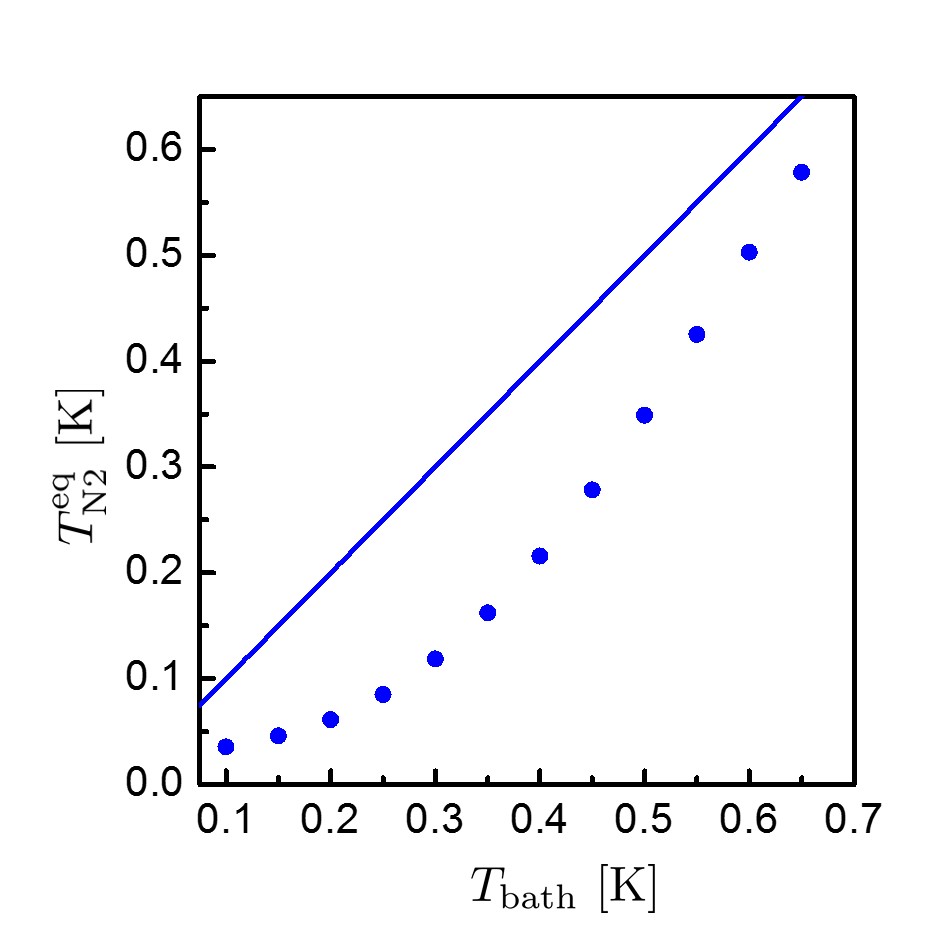}
		\caption{(color online). Steady-state electronic temperature vs Bath temperature. Parameters are $R_2=100\ \Omega$, $T_\textrm{C2}=1.315 \ \textrm K$, $T_\textrm{N2}=T_\textrm{S2}=T_\textrm{bath}$ and $\Gamma_2=10^{-4}\Delta_2$ for the SINIS element. For the thermoelectric we set $P=0.95$, $R_1=1\ \Omega$, $\Gamma_1=10^{-4}\Delta_1$ where $\Delta_1=456\ \mu \textrm{eV}$, $h_{\textrm{exc}}=0.1\Delta_1$ and $Q_\textrm{in}=3\  \textrm{nW}$. The solid line represents the bath temperature $T_\textrm{bath}$.} 
		\label{Fig6}
	\end{centering} 
\end{figure}

In Fig. \ref{Fig6} we show the normal metal steady-state temperature as a function of the bath temperature. We consider a normal island with a volume $\mathcal V_\textrm{N2}=0.01\ \mu\textrm m^3$ and we set a constant input power: $\dot Q_\textrm{in}\sim\ 3\ \textrm{nW}$. The steady state temperature of the normal island is lower than the bath temperature in all the range considered $0.1-0.7$ K. The potential of SINIS refrigeration is well exploited: the NIS energy filtering provides a temperature drop $\sim 200 \textrm{mK}$ at $T_\textrm{bath}=300 \textrm{mK}$, whereas at $T_\textrm{bath}=100 \textrm{mK}$, the electronic temperature is theoretically reduced down to $35\ \textrm{mK}$.
\section{Conclusions}
We have shown that a circuit composed by a N-FI-S thermoelectric element and a SINIS works as a refrigerator for the electronic temperature of the normal layer, under appropriate tuning of the parameters. The system is only biased with heat: the overall mechanism implements therefore a  \textit{cooling by heating} concept.
We analyzed a configuration for the cooler, designed for heating of the superconducting  layer of the N-FI-S element. Based on standard modeling of the N-FI-S and the SINIS element we have established that, with reasonable experimental parameters, the refrigerator can provide cooling power of order $30\ \textrm{pW}$ for an input power of $10\ \textrm{nW}$ and is able to cool the electrons down to $35\ \textrm{mK}$ at $100\ \textrm{mK}$. Its flexibility combined with the simplicity of the design makes it a promising building block for on-chip cooling refrigeration schemes.
\begin{acknowledgments}
We acknowledge the European Research Council under the European Unions Seventh Framework Program (FP7/2007-2013)/ERC Grant agreement No. 615187-COMANCHE and Tuscany Region under the FARFAS 2014 project SCIADRO for partial financial support.
\end{acknowledgments}

%

\begin{thebibliography}{33}%
\makeatletter
\providecommand \@ifxundefined [1]{%
 \@ifx{#1\undefined}
}%
\providecommand \@ifnum [1]{%
 \ifnum #1\expandafter \@firstoftwo
 \else \expandafter \@secondoftwo
 \fi
}%
\providecommand \@ifx [1]{%
 \ifx #1\expandafter \@firstoftwo
 \else \expandafter \@secondoftwo
 \fi
}%
\providecommand \natexlab [1]{#1}%
\providecommand \enquote  [1]{``#1''}%
\providecommand \bibnamefont  [1]{#1}%
\providecommand \bibfnamefont [1]{#1}%
\providecommand \citenamefont [1]{#1}%
\providecommand \href@noop [0]{\@secondoftwo}%
\providecommand \href [0]{\begingroup \@sanitize@url \@href}%
\providecommand \@href[1]{\@@startlink{#1}\@@href}%
\providecommand \@@href[1]{\endgroup#1\@@endlink}%
\providecommand \@sanitize@url [0]{\catcode `\\12\catcode `\$12\catcode
  `\&12\catcode `\#12\catcode `\^12\catcode `\_12\catcode `\%12\relax}%
\providecommand \@@startlink[1]{}%
\providecommand \@@endlink[0]{}%
\providecommand \url  [0]{\begingroup\@sanitize@url \@url }%
\providecommand \@url [1]{\endgroup\@href {#1}{\urlprefix }}%
\providecommand \urlprefix  [0]{URL }%
\providecommand \Eprint [0]{\href }%
\providecommand \doibase [0]{http://dx.doi.org/}%
\providecommand \selectlanguage [0]{\@gobble}%
\providecommand \bibinfo  [0]{\@secondoftwo}%
\providecommand \bibfield  [0]{\@secondoftwo}%
\providecommand \translation [1]{[#1]}%
\providecommand \BibitemOpen [0]{}%
\providecommand \bibitemStop [0]{}%
\providecommand \bibitemNoStop [0]{.\EOS\space}%
\providecommand \EOS [0]{\spacefactor3000\relax}%
\providecommand \BibitemShut  [1]{\csname bibitem#1\endcsname}%
\let\auto@bib@innerbib\@empty
\bibitem [{\citenamefont {Clarke}\ and\ \citenamefont
  {Wilhelm}(2008)}]{Clarke2008}%
  \BibitemOpen
  \bibfield  {author} {\bibinfo {author} {\bibfnamefont {J.}~\bibnamefont
  {Clarke}}\ and\ \bibinfo {author} {\bibfnamefont {F.~K.}\ \bibnamefont
  {Wilhelm}},\ }\href {\doibase 10.1038/nature07128} {\bibfield  {journal}
  {\bibinfo  {journal} {Nature}\ }\textbf {\bibinfo {volume} {453}},\ \bibinfo
  {pages} {1031} (\bibinfo {year} {2008})}\BibitemShut {NoStop}%
\bibitem [{\citenamefont {Devoret}\ and\ \citenamefont
  {Schoelkopf}(2013)}]{Devoret2013}%
  \BibitemOpen
  \bibfield  {author} {\bibinfo {author} {\bibfnamefont {M.~H.}\ \bibnamefont
  {Devoret}}\ and\ \bibinfo {author} {\bibfnamefont {R.~J.}\ \bibnamefont
  {Schoelkopf}},\ }\href {\doibase 10.1126/science.1231930} {\bibfield
  {journal} {\bibinfo  {journal} {Science}\ }\textbf {\bibinfo {volume}
  {339}},\ \bibinfo {pages} {1169} (\bibinfo {year} {2013})}\BibitemShut
  {NoStop}%
\bibitem [{\citenamefont {Giazotto}\ \emph {et~al.}(2006)\citenamefont
  {Giazotto}, \citenamefont {Heikkil{\"{a}}}, \citenamefont {Luukanen},
  \citenamefont {Savin},\ and\ \citenamefont {Pekola}}]{Giazotto2006}%
  \BibitemOpen
  \bibfield  {author} {\bibinfo {author} {\bibfnamefont {F.}~\bibnamefont
  {Giazotto}}, \bibinfo {author} {\bibfnamefont {T.~T.}\ \bibnamefont
  {Heikkil{\"{a}}}}, \bibinfo {author} {\bibfnamefont {A.}~\bibnamefont
  {Luukanen}}, \bibinfo {author} {\bibfnamefont {A.~M.}\ \bibnamefont {Savin}},
  \ and\ \bibinfo {author} {\bibfnamefont {J.~P.}\ \bibnamefont {Pekola}},\
  }\href {\doibase 10.1103/RevModPhys.78.217} {\bibfield  {journal} {\bibinfo
  {journal} {Rev. Mod. Phys.}\ }\textbf {\bibinfo {volume} {78}},\ \bibinfo
  {pages} {217} (\bibinfo {year} {2006})}\BibitemShut {NoStop}%
\bibitem [{\citenamefont {Muhonen}\ \emph {et~al.}(2012)\citenamefont
  {Muhonen}, \citenamefont {Meschke},\ and\ \citenamefont
  {Pekola}}]{Muhonen2012}%
  \BibitemOpen
  \bibfield  {author} {\bibinfo {author} {\bibfnamefont {J.~T.}\ \bibnamefont
  {Muhonen}}, \bibinfo {author} {\bibfnamefont {M.}~\bibnamefont {Meschke}}, \
  and\ \bibinfo {author} {\bibfnamefont {J.~P.}\ \bibnamefont {Pekola}},\
  }\href {\doibase 10.1088/0034-4885/75/4/046501} {\bibfield  {journal}
  {\bibinfo  {journal} {Rep. Prog. Phys.}\ }\textbf {\bibinfo {volume} {75}},\
  \bibinfo {pages} {046501} (\bibinfo {year} {2012})}\BibitemShut {NoStop}%
\bibitem [{\citenamefont {Enss}(2005)}]{Enss2005}%
  \BibitemOpen
  \bibfield  {author} {\bibinfo {author} {\bibfnamefont {C.}~\bibnamefont
  {Enss}},\ }\href@noop {} {\emph {\bibinfo {title} {{Cryogenic particle
  detection}}}}\ (\bibinfo  {publisher} {Springer},\ \bibinfo {year}
  {2005})\BibitemShut {NoStop}%
\bibitem [{\citenamefont {Nahum}\ \emph {et~al.}(1994)\citenamefont {Nahum},
  \citenamefont {Eiles},\ and\ \citenamefont {Martinis}}]{Nahum1994}%
  \BibitemOpen
  \bibfield  {author} {\bibinfo {author} {\bibfnamefont {M.}~\bibnamefont
  {Nahum}}, \bibinfo {author} {\bibfnamefont {T.~M.}\ \bibnamefont {Eiles}}, \
  and\ \bibinfo {author} {\bibfnamefont {J.~M.}\ \bibnamefont {Martinis}},\
  }\href {\doibase 10.1063/1.112456} {\bibfield  {journal} {\bibinfo  {journal}
  {Appl. Phys. Lett.}\ }\textbf {\bibinfo {volume} {65}},\ \bibinfo {pages}
  {3123} (\bibinfo {year} {1994})}\BibitemShut {NoStop}%
\bibitem [{\citenamefont {Bardas}\ and\ \citenamefont
  {Averin}(1995)}]{Bardas1995}%
  \BibitemOpen
  \bibfield  {author} {\bibinfo {author} {\bibfnamefont {A.}~\bibnamefont
  {Bardas}}\ and\ \bibinfo {author} {\bibfnamefont {D.}~\bibnamefont
  {Averin}},\ }\href {\doibase 10.1103/physrevb.52.12873} {\bibfield  {journal}
  {\bibinfo  {journal} {Physical Review B}\ }\textbf {\bibinfo {volume} {52}},\
  \bibinfo {pages} {12873} (\bibinfo {year} {1995})}\BibitemShut {NoStop}%
\bibitem [{\citenamefont {Tinkham}(2004)}]{Tinkham2004}%
  \BibitemOpen
  \bibfield  {author} {\bibinfo {author} {\bibfnamefont {M.}~\bibnamefont
  {Tinkham}},\ }\href@noop {} {\emph {\bibinfo {title} {{Introduction to
  superconductivity}}}}\ (\bibinfo  {publisher} {Dover Publications},\ \bibinfo
  {year} {2004})\BibitemShut {NoStop}%
\bibitem [{\citenamefont {Lowell}\ \emph {et~al.}(2013)\citenamefont {Lowell},
  \citenamefont {O'Neil}, \citenamefont {Underwood},\ and\ \citenamefont
  {Ullom}}]{Lowell2013}%
  \BibitemOpen
  \bibfield  {author} {\bibinfo {author} {\bibfnamefont {P.~J.}\ \bibnamefont
  {Lowell}}, \bibinfo {author} {\bibfnamefont {G.~C.}\ \bibnamefont {O'Neil}},
  \bibinfo {author} {\bibfnamefont {J.~M.}\ \bibnamefont {Underwood}}, \ and\
  \bibinfo {author} {\bibfnamefont {J.~N.}\ \bibnamefont {Ullom}},\ }\href
  {\doibase 10.1063/1.4793515} {\bibfield  {journal} {\bibinfo  {journal}
  {Appl. Phys. Lett.}\ }\textbf {\bibinfo {volume} {102}},\ \bibinfo {pages}
  {082601} (\bibinfo {year} {2013})}\BibitemShut {NoStop}%
\bibitem [{\citenamefont {O'Neil}\ \emph {et~al.}(2012)\citenamefont {O'Neil},
  \citenamefont {Lowell}, \citenamefont {Underwood},\ and\ \citenamefont
  {Ullom}}]{ONeil2012}%
  \BibitemOpen
  \bibfield  {author} {\bibinfo {author} {\bibfnamefont {G.~C.}\ \bibnamefont
  {O'Neil}}, \bibinfo {author} {\bibfnamefont {P.~J.}\ \bibnamefont {Lowell}},
  \bibinfo {author} {\bibfnamefont {J.~M.}\ \bibnamefont {Underwood}}, \ and\
  \bibinfo {author} {\bibfnamefont {J.~N.}\ \bibnamefont {Ullom}},\ }\href
  {\doibase 10.1103/PhysRevB.85.134504} {\bibfield  {journal} {\bibinfo
  {journal} {Phys. Rev. B}\ }\textbf {\bibinfo {volume} {85}},\ \bibinfo
  {pages} {134504} (\bibinfo {year} {2012})}\BibitemShut {NoStop}%
\bibitem [{\citenamefont {Vasenko}\ \emph {et~al.}(2010)\citenamefont
  {Vasenko}, \citenamefont {Bezuglyi}, \citenamefont {Courtois},\ and\
  \citenamefont {Hekking}}]{Vasenko2010}%
  \BibitemOpen
  \bibfield  {author} {\bibinfo {author} {\bibfnamefont {A.~S.}\ \bibnamefont
  {Vasenko}}, \bibinfo {author} {\bibfnamefont {E.~V.}\ \bibnamefont
  {Bezuglyi}}, \bibinfo {author} {\bibfnamefont {H.}~\bibnamefont {Courtois}},
  \ and\ \bibinfo {author} {\bibfnamefont {F.~W.~J.}\ \bibnamefont {Hekking}},\
  }\href {\doibase 10.1103/PhysRevB.81.094513} {\bibfield  {journal} {\bibinfo
  {journal} {Phys. Rev. B}\ }\textbf {\bibinfo {volume} {81}},\ \bibinfo
  {pages} {094513} (\bibinfo {year} {2010})}\BibitemShut {NoStop}%
\bibitem [{\citenamefont {Leivo}\ \emph {et~al.}(1996)\citenamefont {Leivo},
  \citenamefont {Pekola},\ and\ \citenamefont {Averin}}]{Leivo1996}%
  \BibitemOpen
  \bibfield  {author} {\bibinfo {author} {\bibfnamefont {M.~M.}\ \bibnamefont
  {Leivo}}, \bibinfo {author} {\bibfnamefont {J.~P.}\ \bibnamefont {Pekola}}, \
  and\ \bibinfo {author} {\bibfnamefont {D.~V.}\ \bibnamefont {Averin}},\
  }\href {\doibase 10.1063/1.115651} {\bibfield  {journal} {\bibinfo  {journal}
  {Appl. Phys. Lett.}\ }\textbf {\bibinfo {volume} {68}},\ \bibinfo {pages}
  {1996} (\bibinfo {year} {1996})}\BibitemShut {NoStop}%
\bibitem [{\citenamefont {Pekola}\ \emph {et~al.}(2004)\citenamefont {Pekola},
  \citenamefont {Heikkil{\"a}}, \citenamefont {Savin}, \citenamefont
  {Flyktman}, \citenamefont {Giazotto},\ and\ \citenamefont
  {Hekking}}]{Pekola2004}%
  \BibitemOpen
  \bibfield  {author} {\bibinfo {author} {\bibfnamefont {J.~P.}\ \bibnamefont
  {Pekola}}, \bibinfo {author} {\bibfnamefont {T.~T.}\ \bibnamefont
  {Heikkil{\"a}}}, \bibinfo {author} {\bibfnamefont {A.~M.}\ \bibnamefont
  {Savin}}, \bibinfo {author} {\bibfnamefont {J.~T.}\ \bibnamefont {Flyktman}},
  \bibinfo {author} {\bibfnamefont {F.}~\bibnamefont {Giazotto}}, \ and\
  \bibinfo {author} {\bibfnamefont {F.~W.~J.}\ \bibnamefont {Hekking}},\ }\href
  {\doibase 10.1103/PhysRevLett.92.056804} {\bibfield  {journal} {\bibinfo
  {journal} {Phys. Rev. Lett.}\ }\textbf {\bibinfo {volume} {92}},\ \bibinfo
  {pages} {056804} (\bibinfo {year} {2004})}\BibitemShut {NoStop}%
\bibitem [{\citenamefont {Ozaeta}\ \emph {et~al.}(2014)\citenamefont {Ozaeta},
  \citenamefont {Virtanen}, \citenamefont {Bergeret},\ and\ \citenamefont
  {Heikkil{\"a}}}]{Ozaeta2014}%
  \BibitemOpen
  \bibfield  {author} {\bibinfo {author} {\bibfnamefont {A.}~\bibnamefont
  {Ozaeta}}, \bibinfo {author} {\bibfnamefont {P.}~\bibnamefont {Virtanen}},
  \bibinfo {author} {\bibfnamefont {F.~S.}\ \bibnamefont {Bergeret}}, \ and\
  \bibinfo {author} {\bibfnamefont {T.~T.}\ \bibnamefont {Heikkil{\"a}}},\
  }\href {\doibase 10.1103/PhysRevLett.112.057001} {\bibfield  {journal}
  {\bibinfo  {journal} {Phys. Rev. Lett.}\ }\textbf {\bibinfo {volume} {112}},\
  \bibinfo {pages} {057001} (\bibinfo {year} {2014})}\BibitemShut {NoStop}%
\bibitem [{\citenamefont {Machon}\ \emph {et~al.}(2013)\citenamefont {Machon},
  \citenamefont {Eschrig},\ and\ \citenamefont {Belzig}}]{Machon2013}%
  \BibitemOpen
  \bibfield  {author} {\bibinfo {author} {\bibfnamefont {P.}~\bibnamefont
  {Machon}}, \bibinfo {author} {\bibfnamefont {M.}~\bibnamefont {Eschrig}}, \
  and\ \bibinfo {author} {\bibfnamefont {W.}~\bibnamefont {Belzig}},\ }\href
  {\doibase 10.1103/PhysRevLett.110.047002} {\bibfield  {journal} {\bibinfo
  {journal} {Phys. Rev. Lett.}\ }\textbf {\bibinfo {volume} {110}},\ \bibinfo
  {pages} {047002} (\bibinfo {year} {2013})}\BibitemShut {NoStop}%
\bibitem [{\citenamefont {Kolenda}\ \emph {et~al.}(2016)\citenamefont
  {Kolenda}, \citenamefont {Wolf},\ and\ \citenamefont
  {Beckmann}}]{Kolenda2016}%
  \BibitemOpen
  \bibfield  {author} {\bibinfo {author} {\bibfnamefont {S.}~\bibnamefont
  {Kolenda}}, \bibinfo {author} {\bibfnamefont {M.~J.}\ \bibnamefont {Wolf}}, \
  and\ \bibinfo {author} {\bibfnamefont {D.}~\bibnamefont {Beckmann}},\ }\href
  {\doibase 10.1103/PhysRevLett.116.097001} {\bibfield  {journal} {\bibinfo
  {journal} {Phys. Rev. Lett.}\ }\textbf {\bibinfo {volume} {116}},\ \bibinfo
  {pages} {097001} (\bibinfo {year} {2016})}\BibitemShut {NoStop}%
\bibitem [{\citenamefont {Kolenda}\ \emph {et~al.}(2017)\citenamefont
  {Kolenda}, \citenamefont {S\"urgers}, \citenamefont {Fischer},\ and\
  \citenamefont {Beckmann}}]{Kolenda2017}%
  \BibitemOpen
  \bibfield  {author} {\bibinfo {author} {\bibfnamefont {S.}~\bibnamefont
  {Kolenda}}, \bibinfo {author} {\bibfnamefont {C.}~\bibnamefont {S\"urgers}},
  \bibinfo {author} {\bibfnamefont {G.}~\bibnamefont {Fischer}}, \ and\
  \bibinfo {author} {\bibfnamefont {D.}~\bibnamefont {Beckmann}},\ }\href
  {\doibase 10.1103/PhysRevB.95.224505} {\bibfield  {journal} {\bibinfo
  {journal} {Phys. Rev. B}\ }\textbf {\bibinfo {volume} {95}},\ \bibinfo
  {pages} {224505} (\bibinfo {year} {2017})}\BibitemShut {NoStop}%
\bibitem [{\citenamefont {Mari}\ and\ \citenamefont {Eisert}(2012)}]{Mari2012}%
  \BibitemOpen
  \bibfield  {author} {\bibinfo {author} {\bibfnamefont {A.}~\bibnamefont
  {Mari}}\ and\ \bibinfo {author} {\bibfnamefont {J.}~\bibnamefont {Eisert}},\
  }\href {\doibase 10.1103/PhysRevLett.108.120602} {\bibfield  {journal}
  {\bibinfo  {journal} {Phys. Rev. Lett.}\ }\textbf {\bibinfo {volume} {108}},\
  \bibinfo {pages} {120602} (\bibinfo {year} {2012})}\BibitemShut {NoStop}%
\bibitem [{\citenamefont {Cleuren}\ \emph {et~al.}(2012)\citenamefont
  {Cleuren}, \citenamefont {Rutten},\ and\ \citenamefont {Van~den
  Broeck}}]{Clauren2012}%
  \BibitemOpen
  \bibfield  {author} {\bibinfo {author} {\bibfnamefont {B.}~\bibnamefont
  {Cleuren}}, \bibinfo {author} {\bibfnamefont {B.}~\bibnamefont {Rutten}}, \
  and\ \bibinfo {author} {\bibfnamefont {C.}~\bibnamefont {Van~den Broeck}},\
  }\href {\doibase 10.1103/PhysRevLett.108.120603} {\bibfield  {journal}
  {\bibinfo  {journal} {Phys. Rev. Lett.}\ }\textbf {\bibinfo {volume} {108}},\
  \bibinfo {pages} {120603} (\bibinfo {year} {2012})}\BibitemShut {NoStop}%
\bibitem [{\citenamefont {Giazotto}\ \emph {et~al.}(2015)\citenamefont
  {Giazotto}, \citenamefont {Solinas}, \citenamefont {Braggio},\ and\
  \citenamefont {Bergeret}}]{Giazotto2015}%
  \BibitemOpen
  \bibfield  {author} {\bibinfo {author} {\bibfnamefont {F.}~\bibnamefont
  {Giazotto}}, \bibinfo {author} {\bibfnamefont {P.}~\bibnamefont {Solinas}},
  \bibinfo {author} {\bibfnamefont {A.}~\bibnamefont {Braggio}}, \ and\
  \bibinfo {author} {\bibfnamefont {F.~S.}\ \bibnamefont {Bergeret}},\ }\href
  {\doibase 10.1103/PhysRevApplied.4.044016} {\bibfield  {journal} {\bibinfo
  {journal} {Phys. Rev. Appl.}\ }\textbf {\bibinfo {volume} {4}},\ \bibinfo
  {pages} {044016} (\bibinfo {year} {2015})}\BibitemShut {NoStop}%
\bibitem [{\citenamefont {Marchegiani}\ \emph {et~al.}(2016)\citenamefont
  {Marchegiani}, \citenamefont {Virtanen}, \citenamefont {Giazotto},\ and\
  \citenamefont {Campisi}}]{MarchegianiEngine}%
  \BibitemOpen
  \bibfield  {author} {\bibinfo {author} {\bibfnamefont {G.}~\bibnamefont
  {Marchegiani}}, \bibinfo {author} {\bibfnamefont {P.}~\bibnamefont
  {Virtanen}}, \bibinfo {author} {\bibfnamefont {F.}~\bibnamefont {Giazotto}},
  \ and\ \bibinfo {author} {\bibfnamefont {M.}~\bibnamefont {Campisi}},\ }\href
  {\doibase 10.1103/PhysRevApplied.6.054014} {\bibfield  {journal} {\bibinfo
  {journal} {Phys. Rev. Applied}\ }\textbf {\bibinfo {volume} {6}},\ \bibinfo
  {pages} {054014} (\bibinfo {year} {2016})}\BibitemShut {NoStop}%
\bibitem [{\citenamefont {Dynes}\ \emph {et~al.}(1984)\citenamefont {Dynes},
  \citenamefont {Garno}, \citenamefont {Hertel},\ and\ \citenamefont
  {Orlando}}]{Dynes1984}%
  \BibitemOpen
  \bibfield  {author} {\bibinfo {author} {\bibfnamefont {R.~C.}\ \bibnamefont
  {Dynes}}, \bibinfo {author} {\bibfnamefont {J.~P.}\ \bibnamefont {Garno}},
  \bibinfo {author} {\bibfnamefont {G.~B.}\ \bibnamefont {Hertel}}, \ and\
  \bibinfo {author} {\bibfnamefont {T.~P.}\ \bibnamefont {Orlando}},\ }\href
  {\doibase 10.1103/PhysRevLett.53.2437} {\bibfield  {journal} {\bibinfo
  {journal} {Phys. Rev. Lett.}\ }\textbf {\bibinfo {volume} {53}},\ \bibinfo
  {pages} {2437} (\bibinfo {year} {1984})}\BibitemShut {NoStop}%
\bibitem [{\citenamefont {Mermin}(1978)}]{Mermin1978}%
  \BibitemOpen
  \bibfield  {author} {\bibinfo {author} {\bibfnamefont {N.~D.}\ \bibnamefont
  {Mermin}},\ }\href {\doibase 10.1002/piuz.19780090109} {\bibfield  {journal}
  {\bibinfo  {journal} {Phys. unserer Zeit}\ }\textbf {\bibinfo {volume} {9}},\
  \bibinfo {pages} {33} (\bibinfo {year} {1978})}\BibitemShut {NoStop}%
\bibitem [{\citenamefont {Moodera}\ \emph {et~al.}(2007)\citenamefont
  {Moodera}, \citenamefont {Santos},\ and\ \citenamefont
  {Nagahama}}]{Moodera2007}%
  \BibitemOpen
  \bibfield  {author} {\bibinfo {author} {\bibfnamefont {J.~S.}\ \bibnamefont
  {Moodera}}, \bibinfo {author} {\bibfnamefont {T.~S.}\ \bibnamefont {Santos}},
  \ and\ \bibinfo {author} {\bibfnamefont {T.}~\bibnamefont {Nagahama}},\
  }\href {\doibase 10.1088/0953-8984/19/16/165202} {\bibfield  {journal}
  {\bibinfo  {journal} {J. Phys. Condens. Matter}\ }\textbf {\bibinfo {volume}
  {19}},\ \bibinfo {pages} {165202} (\bibinfo {year} {2007})}\BibitemShut
  {NoStop}%
\bibitem [{\citenamefont {Bergeret}\ \emph {et~al.}()\citenamefont {Bergeret},
  \citenamefont {Silaev}, \citenamefont {Virtanen},\ and\ \citenamefont
  {Heikkil{\"a}}}]{Bergeret2017}%
  \BibitemOpen
  \bibfield  {author} {\bibinfo {author} {\bibfnamefont {F.~S.}\ \bibnamefont
  {Bergeret}}, \bibinfo {author} {\bibfnamefont {M.}~\bibnamefont {Silaev}},
  \bibinfo {author} {\bibfnamefont {P.}~\bibnamefont {Virtanen}}, \ and\
  \bibinfo {author} {\bibfnamefont {T.~T.}\ \bibnamefont {Heikkil{\"a}}},\
  }\href@noop {} {\ }\Eprint {http://arxiv.org/abs/arXiv:1706.08245}
  {arXiv:1706.08245} \BibitemShut {NoStop}%
\bibitem [{\citenamefont {Moodera}\ \emph {et~al.}(1988)\citenamefont
  {Moodera}, \citenamefont {Hao}, \citenamefont {Gibson},\ and\ \citenamefont
  {Meservey}}]{Moodera1988}%
  \BibitemOpen
  \bibfield  {author} {\bibinfo {author} {\bibfnamefont {J.~S.}\ \bibnamefont
  {Moodera}}, \bibinfo {author} {\bibfnamefont {X.}~\bibnamefont {Hao}},
  \bibinfo {author} {\bibfnamefont {G.~A.}\ \bibnamefont {Gibson}}, \ and\
  \bibinfo {author} {\bibfnamefont {R.}~\bibnamefont {Meservey}},\ }\href
  {\doibase 10.1103/PhysRevLett.61.637} {\bibfield  {journal} {\bibinfo
  {journal} {Phys. Rev. Lett.}\ }\textbf {\bibinfo {volume} {61}},\ \bibinfo
  {pages} {637} (\bibinfo {year} {1988})}\BibitemShut {NoStop}%
\bibitem [{\citenamefont {Moodera}\ \emph {et~al.}(1993)\citenamefont
  {Moodera}, \citenamefont {Meservey},\ and\ \citenamefont
  {Hao}}]{Moodera1993}%
  \BibitemOpen
  \bibfield  {author} {\bibinfo {author} {\bibfnamefont {J.~S.}\ \bibnamefont
  {Moodera}}, \bibinfo {author} {\bibfnamefont {R.}~\bibnamefont {Meservey}}, \
  and\ \bibinfo {author} {\bibfnamefont {X.}~\bibnamefont {Hao}},\ }\href
  {\doibase 10.1103/PhysRevLett.70.853} {\bibfield  {journal} {\bibinfo
  {journal} {Phys. Rev. Lett.}\ }\textbf {\bibinfo {volume} {70}},\ \bibinfo
  {pages} {853} (\bibinfo {year} {1993})}\BibitemShut {NoStop}%
\bibitem [{\citenamefont {Hao}\ \emph {et~al.}(1990)\citenamefont {Hao},
  \citenamefont {Moodera},\ and\ \citenamefont {Meservey}}]{Hao1990}%
  \BibitemOpen
  \bibfield  {author} {\bibinfo {author} {\bibfnamefont {X.}~\bibnamefont
  {Hao}}, \bibinfo {author} {\bibfnamefont {J.~S.}\ \bibnamefont {Moodera}}, \
  and\ \bibinfo {author} {\bibfnamefont {R.}~\bibnamefont {Meservey}},\ }\href
  {\doibase 10.1103/PhysRevB.42.8235} {\bibfield  {journal} {\bibinfo
  {journal} {Phys. Rev. B}\ }\textbf {\bibinfo {volume} {42}},\ \bibinfo
  {pages} {8235} (\bibinfo {year} {1990})}\BibitemShut {NoStop}%
\bibitem [{\citenamefont {Strambini}\ \emph {et~al.}(2017)\citenamefont
  {Strambini}, \citenamefont {Golovach}, \citenamefont {De~Simoni},
  \citenamefont {Moodera}, \citenamefont {Bergeret},\ and\ \citenamefont
  {Giazotto}}]{Strambini2017}%
  \BibitemOpen
  \bibfield  {author} {\bibinfo {author} {\bibfnamefont {E.}~\bibnamefont
  {Strambini}}, \bibinfo {author} {\bibfnamefont {V.~N.}\ \bibnamefont
  {Golovach}}, \bibinfo {author} {\bibfnamefont {G.}~\bibnamefont {De~Simoni}},
  \bibinfo {author} {\bibfnamefont {J.~S.}\ \bibnamefont {Moodera}}, \bibinfo
  {author} {\bibfnamefont {F.~S.}\ \bibnamefont {Bergeret}}, \ and\ \bibinfo
  {author} {\bibfnamefont {F.}~\bibnamefont {Giazotto}},\ }\href {\doibase
  10.1103/PhysRevMaterials.1.054402} {\bibfield  {journal} {\bibinfo  {journal}
  {Phys. Rev. Materials}\ }\textbf {\bibinfo {volume} {1}},\ \bibinfo {pages}
  {054402} (\bibinfo {year} {2017})}\BibitemShut {NoStop}%
\bibitem [{\citenamefont {Benenti}\ \emph {et~al.}(2017)\citenamefont
  {Benenti}, \citenamefont {Casati}, \citenamefont {Saito},\ and\ \citenamefont
  {Whitney}}]{Benenti2016}%
  \BibitemOpen
  \bibfield  {author} {\bibinfo {author} {\bibfnamefont {G.}~\bibnamefont
  {Benenti}}, \bibinfo {author} {\bibfnamefont {G.}~\bibnamefont {Casati}},
  \bibinfo {author} {\bibfnamefont {K.}~\bibnamefont {Saito}}, \ and\ \bibinfo
  {author} {\bibfnamefont {R.}~\bibnamefont {Whitney}},\ }\href {\doibase
  https://doi.org/10.1016/j.physrep.2017.05.008} {\bibfield  {journal}
  {\bibinfo  {journal} {Physics Reports}\ }\textbf {\bibinfo {volume} {694}},\
  \bibinfo {pages} {1 } (\bibinfo {year} {2017})}\BibitemShut {NoStop}%
\bibitem [{\citenamefont {Agulo}\ \emph {et~al.}(2004)\citenamefont {Agulo},
  \citenamefont {Kuzmin}, \citenamefont {Fominsky},\ and\ \citenamefont
  {Tarasov}}]{Agulo2004}%
  \BibitemOpen
  \bibfield  {author} {\bibinfo {author} {\bibfnamefont {I.~J.}\ \bibnamefont
  {Agulo}}, \bibinfo {author} {\bibfnamefont {L.}~\bibnamefont {Kuzmin}},
  \bibinfo {author} {\bibfnamefont {M.}~\bibnamefont {Fominsky}}, \ and\
  \bibinfo {author} {\bibfnamefont {M.}~\bibnamefont {Tarasov}},\ }\href
  {\doibase 10.1088/0957-4484/15/4/020} {\bibfield  {journal} {\bibinfo
  {journal} {Nanotechnology}\ }\textbf {\bibinfo {volume} {15}},\ \bibinfo
  {pages} {S224} (\bibinfo {year} {2004})}\BibitemShut {NoStop}%
\bibitem [{\citenamefont {Rajauria}\ \emph {et~al.}(2012)\citenamefont
  {Rajauria}, \citenamefont {Pascal}, \citenamefont {Gandit}, \citenamefont
  {Hekking}, \citenamefont {Pannetier},\ and\ \citenamefont
  {Courtois}}]{Rajauria2012}%
  \BibitemOpen
  \bibfield  {author} {\bibinfo {author} {\bibfnamefont {S.}~\bibnamefont
  {Rajauria}}, \bibinfo {author} {\bibfnamefont {L.~M.~A.}\ \bibnamefont
  {Pascal}}, \bibinfo {author} {\bibfnamefont {P.}~\bibnamefont {Gandit}},
  \bibinfo {author} {\bibfnamefont {F.~W.~J.}\ \bibnamefont {Hekking}},
  \bibinfo {author} {\bibfnamefont {B.}~\bibnamefont {Pannetier}}, \ and\
  \bibinfo {author} {\bibfnamefont {H.}~\bibnamefont {Courtois}},\ }\href
  {\doibase 10.1103/PhysRevB.85.020505} {\bibfield  {journal} {\bibinfo
  {journal} {Phys. Rev. B}\ }\textbf {\bibinfo {volume} {85}},\ \bibinfo
  {pages} {020505} (\bibinfo {year} {2012})}\BibitemShut {NoStop}%
\bibitem [{\citenamefont {Court}\ \emph {et~al.}(2008)\citenamefont {Court},
  \citenamefont {Ferguson}, \citenamefont {Lutchyn},\ and\ \citenamefont
  {Clark}}]{Court2008}%
  \BibitemOpen
  \bibfield  {author} {\bibinfo {author} {\bibfnamefont {N.~A.}\ \bibnamefont
  {Court}}, \bibinfo {author} {\bibfnamefont {A.~J.}\ \bibnamefont {Ferguson}},
  \bibinfo {author} {\bibfnamefont {R.}~\bibnamefont {Lutchyn}}, \ and\
  \bibinfo {author} {\bibfnamefont {R.~G.}\ \bibnamefont {Clark}},\ }\href
  {\doibase 10.1103/PhysRevB.77.100501} {\bibfield  {journal} {\bibinfo
  {journal} {Phys. Rev. B}\ }\textbf {\bibinfo {volume} {77}},\ \bibinfo
  {pages} {100501} (\bibinfo {year} {2008})}\BibitemShut {NoStop}%
\end{thebibliography}
\end{document}